# DISTRIBUTION OF SUPERNOVAE IN SPIRAL GALAXIES

Sidney van den Bergh

Dominion Astrophysical Observatory, National Research Council

5071 West Saanich Road, Victoria, British Columbia

V8X 4M6, Canada

Electronic mail: sidney.vandenbergh@hia.nrc.ca

**ABSTRACT**

Data are presently available on 156 supernovae (SNe) of known type that have occurred in spirals of types Sa-Sc having inclinations $i < 70°$. The surface density distribution of SN Ibc + SN II is found to be an exponential with exponent $-4.5\, r_{SN} / R_{25}$. Available data suggest, but do not yet prove, that SNe Ibc may be more centrally concentrated than SNe II. It is tentatively suggested that such an effect might be due to preferential mass loss via stellar winds, in the relatively metal-rich progenitors of core-collapse supernovae (or by an enhanced mass transfer rate in binary SN progenitors) in the high metallicity central regions of spiral galaxies.

## 1. INTRODUCTION

The overall frequency distribution of supernovae in galaxies of differing types has been discussed by van den Bergh (1960), Bertaud (1961), Minkowski (1964), McCarthy (1973), Iye & Kodeira (1975), Oemler & Tinsley (1979) and Cappellaro et al. (1993). For a review of this subject the reader is referred to van den Bergh & Tammann (1991) and van den Bergh (1991). The global distribution of supernovae within galaxies was first discussed by Whipple (1939) who found $\sigma_{SN} \sim r_{SN}^{-1}$, and by Johnson & MacLeod (1963) who noted a possible deficiency of SNe in the (over-exposed) central regions of galaxies. For a detailed recent discussion of the global distribution of SNe in galaxies the reader is referred to Barnutov, Makarova & Tsvetkov (1992) and references therein. Possible differences in the radial distributions of supernovae in normal and active galaxies have been studied by Turatto et al. (1989) and Petrosian & Turatto (1990). The apparent association of SNe II with spiral arms was first noted by Reaves (1953) and has subsequently been studied in more detail by Johnson & MacLeod and by Maza & van den Bergh (1976). More recently, the distribution of SNe relative to spiral arms and H II regions has been investigated by Van Dyk (1992), by Barnutov, Tsvetkov & Filimonova (1994), by Van Dyk, Hamuy & Filippenko (1996), by Barth et al. (1996) and by McMillan & Ciardullo (1997).

The results of these investigations may be summarized as follows:

1.  Only SNe Ia have been observed to occur in elliptical galaxies.

2.  Supernovae of Types Ia, Ibc and II are all observed to occur in spiral galaxies. In early-type spirals the frequency of SNe Ia per unit luminosity is higher than that of SNe II. However, SNe II have a higher specific frequency in late-type spirals than do SNe Ia.

3.  Although statistics are still poor, it appears that the intrinsic frequency of SNe Ia might be very high in objects such as the Am galaxy NGC 5253, that seem to have undergone episodes of violent star formation ~ 1 Gyr ago (van den Bergh 1980). Radio observations of M82 (Kronberg, Biermann & Schwab 1985) suggest that this galaxy (which is presently undergoing a burst of star formation) is producing SNe (probably of Types Ibc and II) at a high rate.

4.  SNe II are found to be more concentrated to spiral arms than are SNe Ia (Maza & van den Bergh 1976), but both types occur closer to arms than a random disk population (McMillan & Ciardullo 1997).

5.  The degree to which SNe Ibc and SNe II are associated with H II regions is similar. This suggests that both SNe Ibc and SNe II have progenitors in the same mass range (Barnutov, Tsvetkov & Filimonova 1994, Van Dyk, Hamuy & Filippenko 1996).

The present paper differs from that of Barnutov et al. (1992) in three respects: (1) Only supernovae with spectroscopically determined types Ia, Ibc or II in spiral galaxies of known radial velocity having i < 70° are included, (2) SNe with classifications made prior to 1985, that were marked as being uncertain, were omitted, and (3) the cut-off date of the database is advanced from 1991 October 1 to 1996 June 15, resulting in a gain of 47 supernovae (18 SNe Ia, 4 SNe Ibc, 25 SNe II).

**2.     SUPERNOVA DATABASE**

The present investigation was based on the electronic supernova catalog of Barnutov, Tsvetkov & Filimonova (available at http://www.sai.msu.su/group/sn). A check on the quoted supernova types was obtained by comparison with the types listed by Cappellaro et al., which are available at http://www.pd.astro.it/supern/.

The only substantive difference between supernova classifications in these two catalogs were the following: SN 1961V and SN 1964A are listed as being of Type IIp in the Russian catalog and as Type V in the Italian catalog. Also, SN 1971G is listed as being of Type Ia in the Russian catalog and as Type I in the Italian catalog. In all three cases the listing in the Russian catalog was adopted.

There were also objects classified as Ib in one catalog, and as Ic in the other. In the present investigation SNe Ib and SNe Ic were all lumped together as Type Ibc.

The distinction between supernovae of Types Ia, Ib and Ic was not understood until 1985 (Elias et al. 1985, Panagia 1985, Uomoto & Kirshner 1985, Wheeler & Levreault 1985). The sample was therefore sub-divided into sample A, which contained all SNe of Types Ia, Ib, Ic and II that occurred prior to 1985. [Objects classified as I, or for which the classification was uncertain (:), were excluded from sample A]. Sample B contained <u>all</u> observations of SNe that reached maximum after 1985 January 1. SN 1993R, classified only as Type Ipec, was omitted from sample B. Data in sample B, which are based on more recent observations and classifications should, in general, be more reliable than those in sample A. However, some of the old data are based on recent re-interpretations of archival spectra. The last supernova included in the present investigation was SN 1996aj, which reached maximum on 1996 June 15. Only supernovae for which radial velocities were available and that occurred in spiral galaxies with types T1 to T6 (eg. de Vaucouleurs, de Vaucouleurs & Corwin 1976), corresponding to Hubble types Sa-Sc, were included in the present database. Furthermore, to minimize absorption and projection effects, spirals with $i \geq 70°$ were excluded from the present study. The galaxy radius $R_{25}$ used in the present investigation is the apparent semi-major axes of the supernova parent galaxy isophote having $\mu_B = 25.0$ mag arcsec$^{-2}$. The (projected) linear distance $r_{SN}$ of individual supernovae from the nuclei of their parent galaxies were computed by assuming $H_o = 75$ km s$^{-1}$ Mpc$^{-1}$. A distance $D = 16.0$ Mpc was assumed for galaxies in the Virgo cluster. Distances to a few nearby galaxies were taken from a variety of recent sources. The total number of supernovae in sample A [T(max) < 1985.0] was 62, of which 17 were of Type Ia, 6 of Type Ibc and 39 of Type II. Sample B [T(max) > 1985.0] comprised 94 supernovae, of which 37 were of Type Ia, 12 of Type Ib and 45 of Type II.

**3.    DISCUSSION**

Figures 1-3 exhibit plots of $r_{SN} / R_{25}$ versus log V in km s$^{-1}$ for supernovae of Types II, Ibc and Ia, respectively. As expected, all three figures show that supernovae which were observed long ago (sample A = open circles), on average, had lower radial velocities than supernovae observed more recently (sample B = filled circles). Inspection of Fig. 1 also suggests that there is a deficiency of SNe II with $r_{SN} / R_{25} \sim 0.2$ for log V > 3.5. This is, no doubt, due to the Shaw (1979) effect, i.e. the failure to discover some supernovae in the high surface brightness central regions of distant galaxies. Barnutov et al. (1992) and Cappellaro et al. (1993) show that such radial selection effects differ somewhat for different supernova search programs. In the following it will (conservatively) be assumed

that the Shaw effect may be neglected for SNe Ibc and SNe II out to log V = 3.4, and for the (typically more luminous) SNe Ia out to log V = 3.6.

The distribution of $r_{SN} / R_{25}$ for SNe Ia with log V < 3.6 and for SNe Ibc and SNe II with log V < 3.4 is given in Table 1 and plotted in Fig. 4. This figure suggests that <u>SNe Ibc might be more strongly concentrated to the nuclei of their parent galaxies than is the case for SNe II</u>. A Kolmogorov-Smirnov test shows that there is only a 12% probability that the ancestors of SNe Ibc and of SNe II had the same radial density distribution. However, more observations will be required to establish (or disprove) this conclusion at a respectable level of statistical confidence. Six out of 16 SNe Ibc (38%) are found to be located within 2.0 kpc of the nuclei of their parent galaxies, compared to 10 out of 58 (17%) for all SNe II, one out of 11 (9%) of SNe II L, and only three out of 37 (8%) for SNe Ia. If real, the apparent difference between the radial distributions of SNe Ibc and SNe II might be a consequence of radial abundance gradients in spirals. Such radial abundance gradients could either affect the mass-loss rate of single SNe directly, or it could influence the rate of mass transfer in binary supernova progenitors (Podsialowski, Joss & Hsu 1992). Stronger mass loss in stellar winds of massive metal-rich supernovae in the cores of spirals might (Kudritzki, Pauldrach & Puls 1987) might result in more mass loss than in the winds of more metal-deficient massive stars in the outer disks of spirals. Specifically (Abbott,

1982, Kudritzki 1996) one expects

$$\dot{M}(Z) \approx \dot{M}(\odot) [Z/Z(\odot)]^{\alpha}, \qquad (1)$$

in which $\alpha$ is of order 0.5 to 1.0. As a result of such selective mass loss massive stars near galactic nuclei might have a higher probability of losing their outer hydrogen envelopes before exploding, than is the case for relatively metal-poor massive supernova progenitors in the more metal-deficient outer disks of spirals.

Barnutov et al. (1994) and Van Dyk et al. (1996) have shown that SNe Ibc and SNe II are similarly associated with massive stars and H II regions. This suggests that they have progenitors of comparable mass. The radial distributions of SNe Ibc and SNe II have therefore been studied together. Since only 22% of the SNe Ibc + SNe II in Table 1 are of type Ibc the derived radial distribution of SNe Ibc + SNe II is dominated by the SNe II.

The data on core-collapsed supernovae (SNe Ibc + SN II) from Table 1 have been used to calculate the surface density (in arbitrary units) of such young supernovae as a function of $r_{SN} / R_{25}$. Figure 5 shows that the data are well represented by a single relation of the form

$$\sigma_{SN} = \sigma_o \exp(-4.5\, r_{SN} / R_{25}). \qquad (2)$$

Barnutov et al. (1992) point out that the <u>stellar</u> surface density gradient in spiral galaxies is steeper for $r_{SN} / R_{25} < 0.5$ than it is for $r_{SN} / R_{25} > 0.5$. However, this difference is probably mainly due to the presence of old central bulges which do not contribute to the population of massive core-collapse supernovae. Because the SNe in the present sample occurred in mixed population spirals having types Sa, Sb and Sc (which have a wide range of disk characteristics) it is not possible to relate Eqn. (2) directly to the characteristics of the young disk populations in spiral galaxies.

Figure 6 shows that the number of SNe Ia observed to date is too small to constrain the slope of the $\sigma_{SN}$ (Ia) versus $r_{SN} / R_{25}$ relation very well. However, the data plotted in Fig. 6 show that a relation of the form

$$\sigma_{SN} \approx \sigma_o \exp(-5.5 \, r_{SN} / R_{25}) \qquad (3)$$

provides an acceptable fit to the rather scanty observational data that are presently available. The fact that Eqn. (3) appears to exhibit a steeper radial fall-off than Eqn. (2) indicates that the old population of SNe Ia progenitors is more centrally concentrated than the disk-dominated distributions of SN Ibc + SN II progenitors.

It is a pleasure to thank Daniel Durand for his assistance with the Italian and Russian supernovae databases. I also thank Alex Filippenko for helpful comments on selection effects and their possible influence on supernova discoveries and David Branch for comments on the classification of SNe Ibc. Finally, I am indebted to Rolf Kudritzki for information on the effect of metallicity on the mass-loss rates from O-type stars and to a particularly helpful referee.

**Table 1. Distribution of Supernovae**

| $r_{SN} / R_{25}$ | Ia | Ibc | II |
|---|---|---|---|
| 0.00 - 0.09 | 0 | 4 | 3 |
| 0.10  0.19 | 8 | 4 | 6 |
| 0.20  0.29 | 8 | 2 | 11 |
| 0.30  0.39 | 10 | 1 | 6 |
| 0.40  0.49 | 3 | 3 | 9 |
| 0.50  0.59 | 1 | 1 | 7 |
| 0.60  0.69 | 3 | 1 | 5 |
| 0.70  0.79 | 3 | 0 | 3 |
| 0.80  0.89 | 0 | 0 | 4 |
| 0.90  0.99 | 0 | 0 | 0 |
| 1.00  1.09 | 0 | 0 | 2 |
| 1.10  1.19 | 0 | 0 | 0 |
| $\geq 1.20$ | 1 | 0 | 2 |

**Figure Legends**

Fig. 1  Log V versus $r_{SN}$ for supernovae of Type II.  Sample A [T(max) < 1985.0] dots (•) and Sample B [T(max) > 1985.0] plus (+) signs.  The figure shows that (1) older supernovae are, on average, closer than ones that have been discovered more recently, and (2) a deficiency of SNe II at small values of $r_{SN} / R_{25}$ among distant objects with redshifts > 3.5 dex, which is due to the Shaw effect.

Fig. 2  Log V versus $r_{SN} / R_{25}$ for SNe Ibc.  The figure suggests the possibility that SNe Ibc might be more concentrated to galaxy nuclei than SNe II.  Symbol coding as in Fig. 1.

Fig. 3  Log V versus $r_{SN} / R_{25}$ for SNe Ia.  Data are coded as in Fig. 1.

Fig. 4  Radial distribution of supernovae of different types.  This figures suggests that SNe Ibc might be more concentrated towards the nuclei of their parent galaxies than is the case for SNe II.

Fig. 5  Surface density of supernovae of Types Ibc and II (on arbitrary scale), versus $r_{SN} / R_{25}$. The data are seen to be well-represented by the relation given in Eqn. (2).

Fig. 6  Surface density of SNe Ia as a function of $r_{SN} / R_{25}$. The line shown in the figure is given by Eqn. (3).

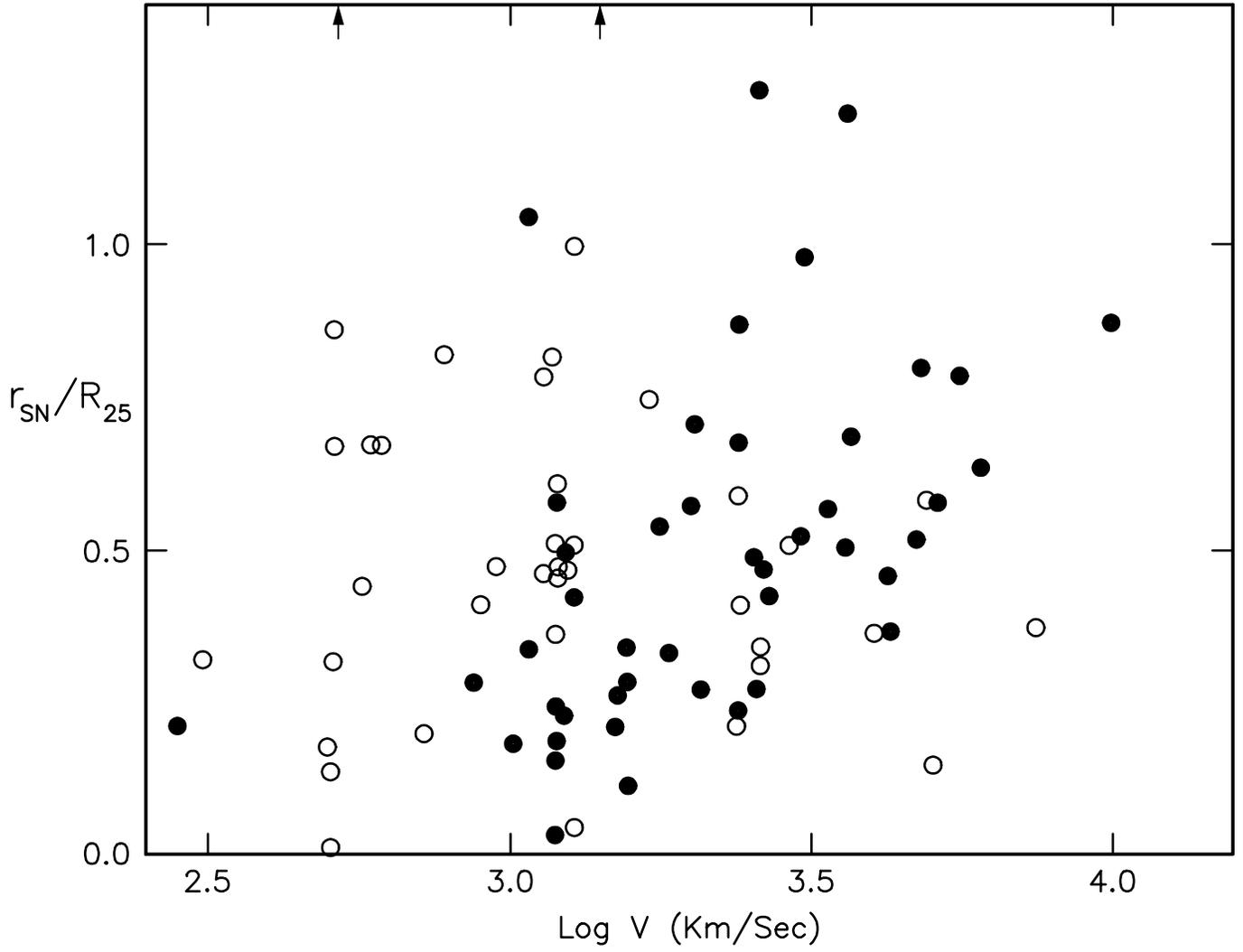

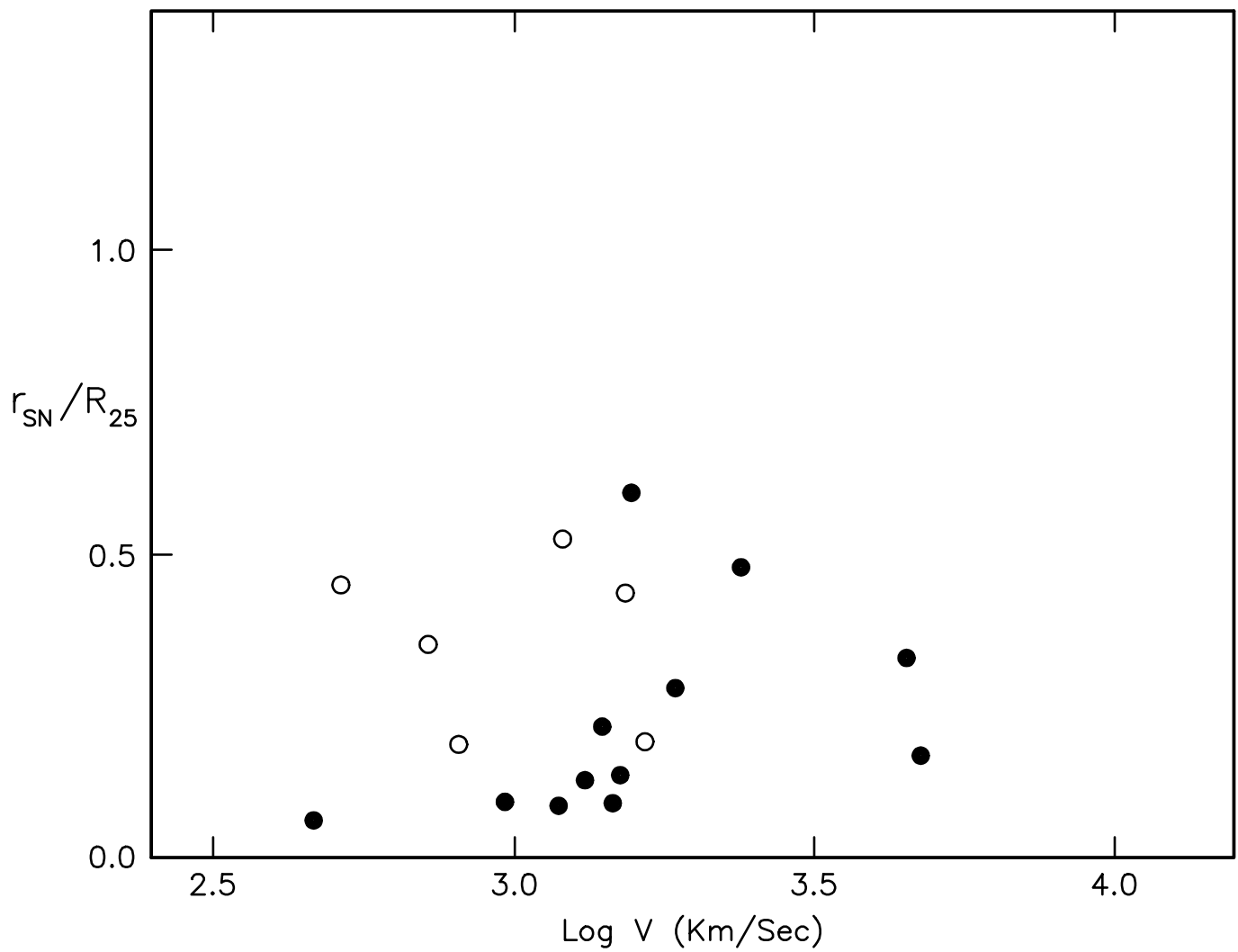

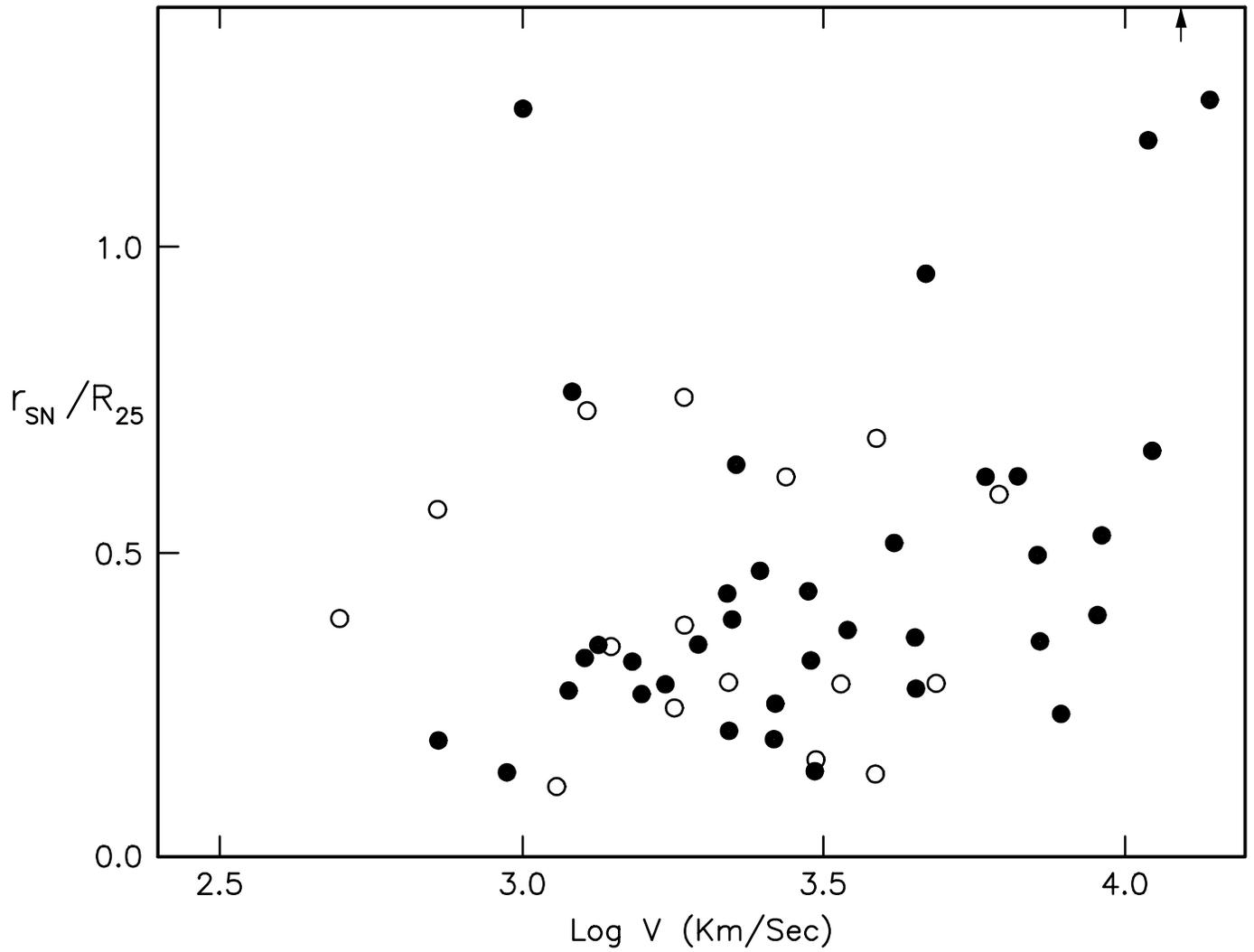

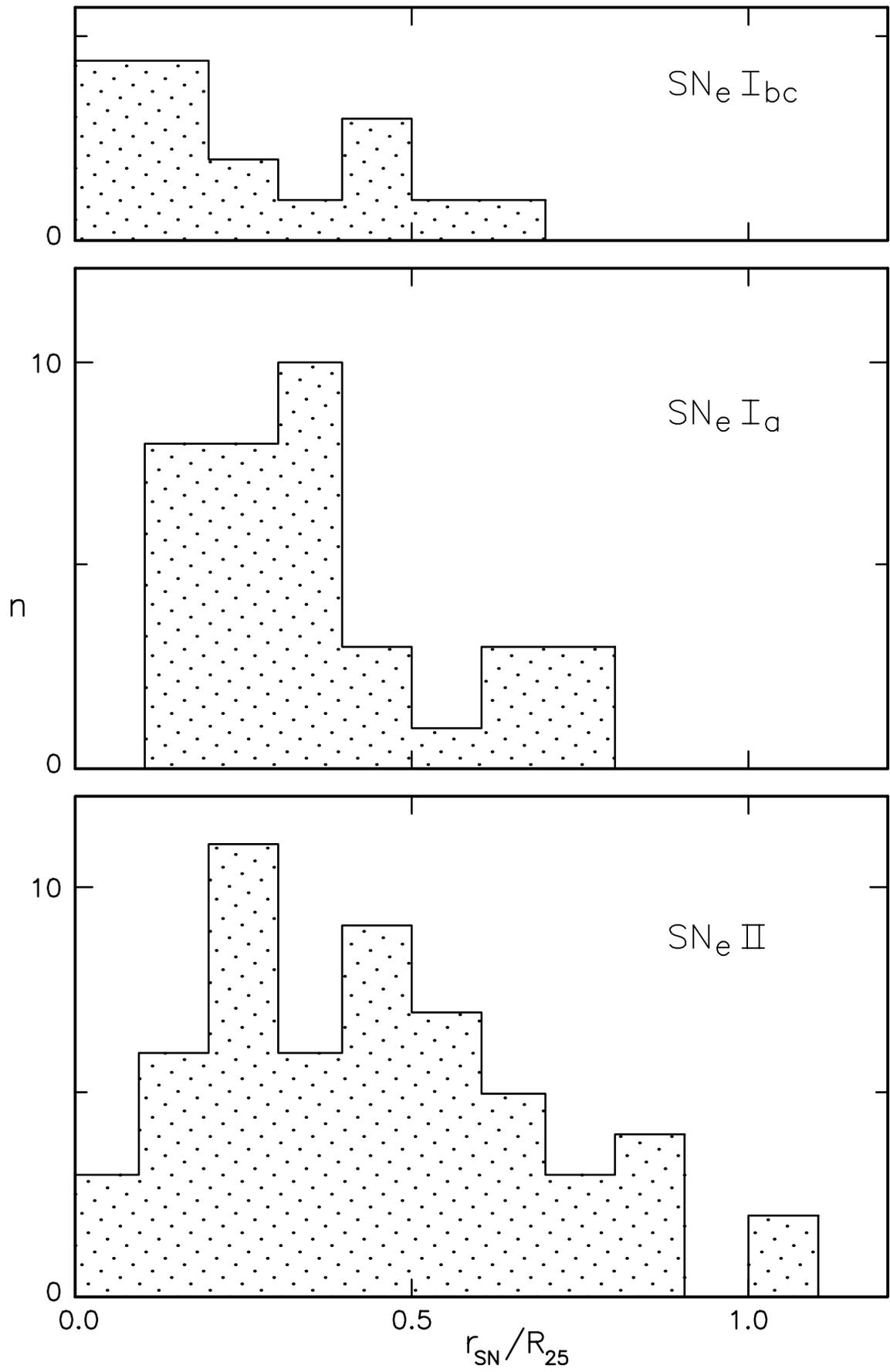

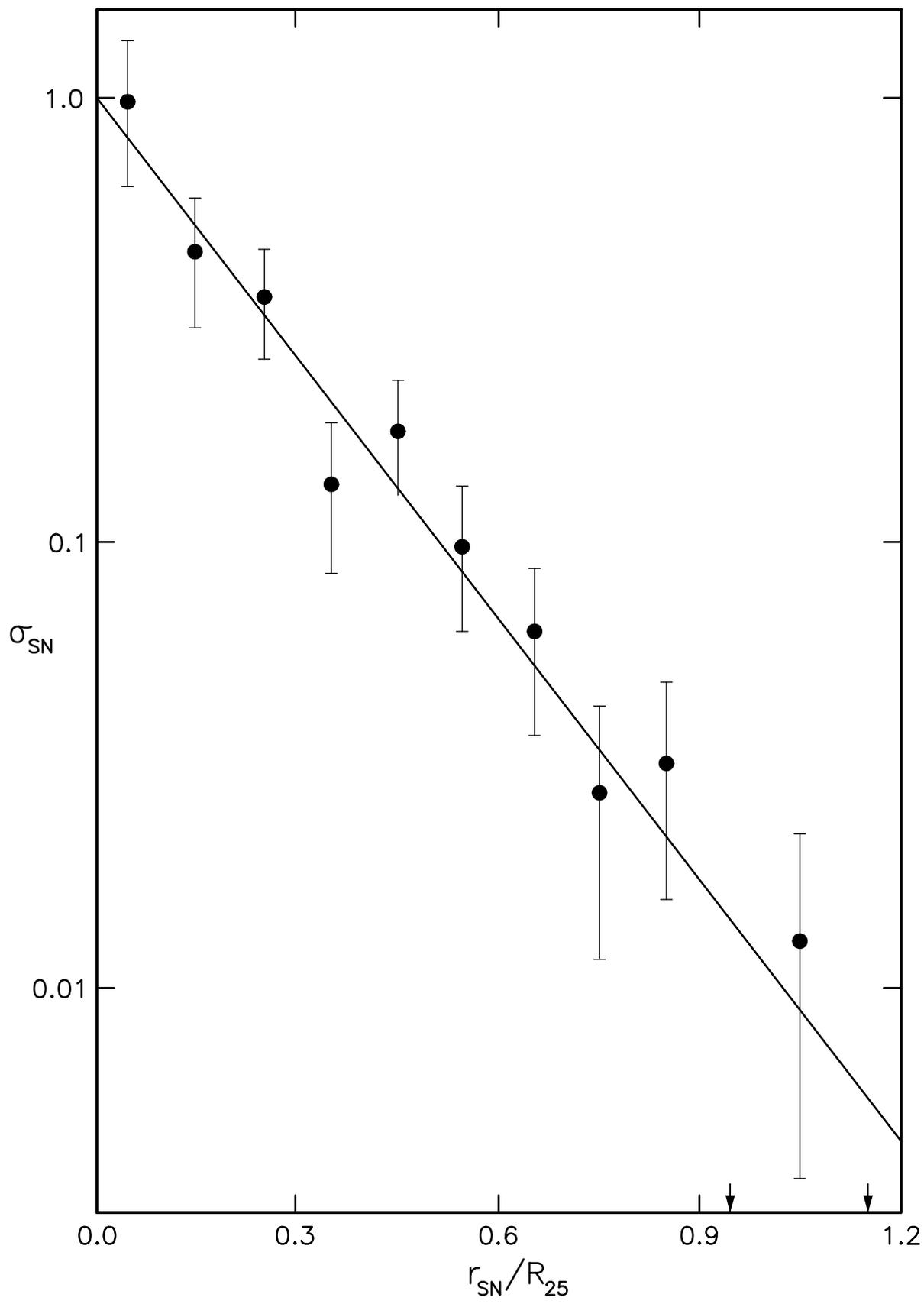

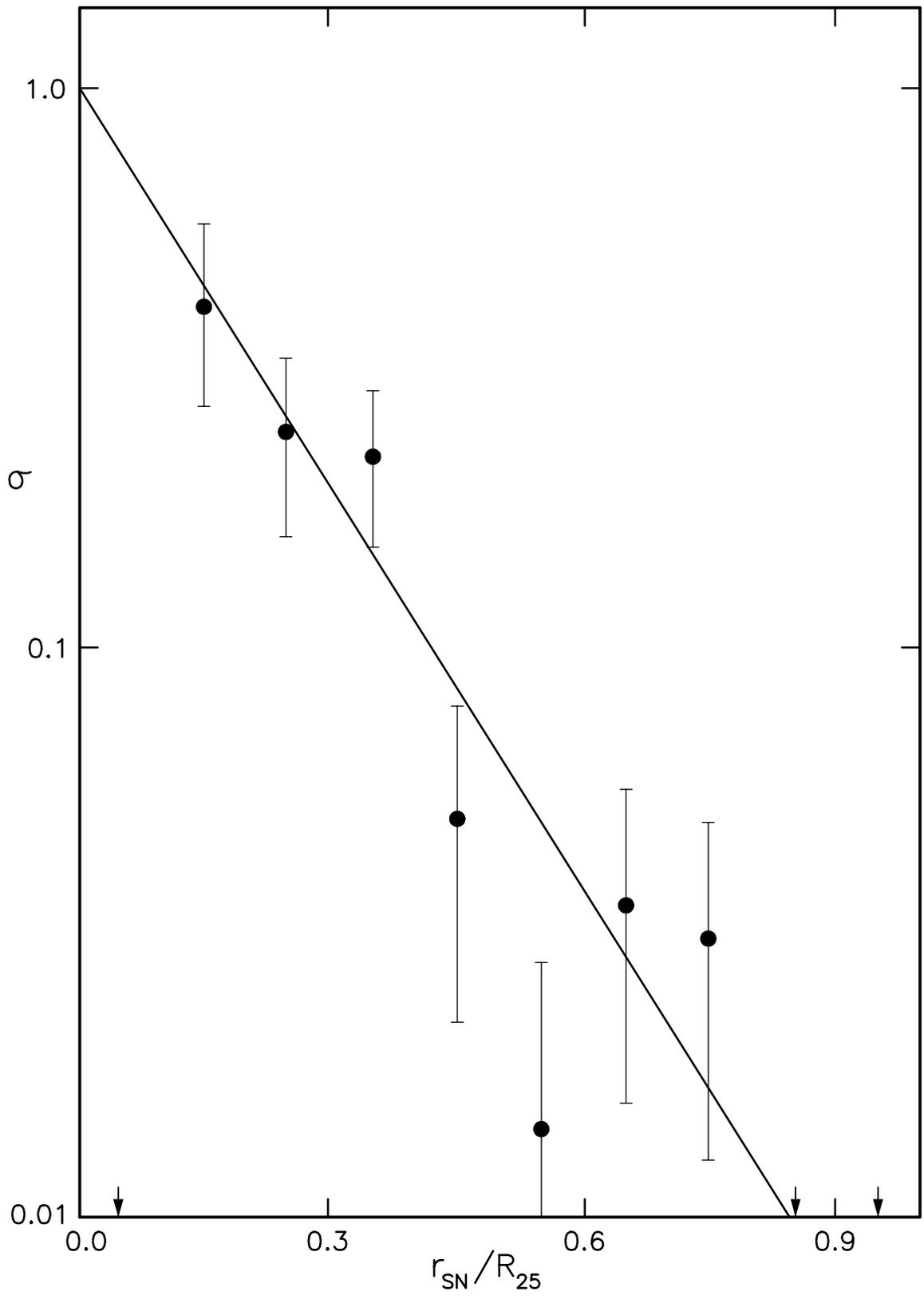